\documentclass[twocolumn]{aastex63}

\graphicspath{{./}{figures/}}

\received{January 1, 2018}
\revised{January 7, 2018}
\accepted{\today}

\submitjournal{ApJ}

\shorttitle{Observation of S62 in 2019}
\shortauthors{Pei$\beta$ker et al.}

\begin{document}

\title{Observation of the apoapsis of S62 in 2019 with NIRC2 and SINFONI}

\correspondingauthor{Florian Pei{\ss}ker}
\email{peissker@ph1.uni-koeln.de}
\author[0000-0002-9850-2708]{Florian Pei$\beta$ker}
\affil{I.Physikalisches Institut der Universit\"at zu K\"oln, Z\"ulpicher Str. 77, 50937 K\"oln, Germany}
\author{Andreas Eckart}
\affil{I.Physikalisches Institut der Universit\"at zu K\"oln, Z\"ulpicher Str. 77, 50937 K\"oln, Germany}
\affil{Max-Plank-Institut f\"ur Radioastronomie, Auf dem H\"ugel 69, 53121 Bonn, Germany}
\author{Basel Ali}
\affil{I.Physikalisches Institut der Universit\"at zu K\"oln, Z\"ulpicher Str. 77, 50937 K\"oln, Germany}

\begin{abstract}

Given the increased attention towards the detection of faint stars in the Galactic center, we would like to address the detectability of S62 in its apoapsis with SINFONI (VLT) and NIRC2 (KECK) in this work. Because of the nearby stars and the chance of confusion, we are using Lucy-Richardson deconvolved images to detect S62 on its Keplerian orbit around Sgr~A* with a period of less than 10 years. We use the same dataset as for S62 to trace additionally the S-cluster star S29 at the expected position based on the orbital elements presented in this work. 
To verify the results of the filtering technique, we are analysing K-band continuum data of the same epoch independently observed with NIRC2/KECK. Based on the well-derived orbit of S62, we find the star in projection at the expected position in 2019.4 and 2019.5. By analyzing the SINFONI data of 2019.5, we confirm the $16.1\,\pm\,0.2$ mag for S62 that was formerly derived with NACO (VLT). We base our NACO imaging analysis on the robust data set that was previously used to investigate the Schwarzschild precision of S2. We also present a critical discussion of the elsewhere proposed linear trajectory of S62 and its disputed identification with a 19 mag star found with GRAVITY mounted at the VLT Interferometer.

\end{abstract}


\section{Introduction} \label{sec:intro}

For the first time, \cite{Eckart2002} showed that the S-star S2 \citep[][]{Schoedel2002} is on a bound orbit around a compact mass of about $4\,\times\,10^6\,M_{\odot}$ at the position of the prominent radio source Sgr~A*.
\normalfont{Consequently, the intense monitoring of S2 over the last 30 years including its periapse helped to understand the imprint of gravitational effects on its orbit \citep[][]{Parsa2017, gravity2018, Do2019_S2}. In addition, near-infrared observations of the star or the surrounding ionized gas helped to constrain accretion models \citep[e.g.,][]{Hosseini2020, ciurlo2021}. Using a multiwavelength approach by expanding theses observations to the submillimeter and 2-8 keV resulted in a comprehensive description of the variability of Sgr~A* \citep[][]{Witzel2020}. Furthermore, analysing the lightcurve of flares from Sgr~A* can result in an estimate of the black hole mass that is also applicable to extragalactic sources \citep[][]{Karssen2017}.\newline From an observational point of view, the first light of the IR instrument GRAVITY\footnote{Installed at the Very Large Telescope Interferometer (VLTI).} \citep[][]{GravityCollaboration2017} increased the capability of detecting faint objects \citep[][]{GravityCollaboration2021}. In contrast, using publically available data that cover almost 20 years in combination with enhanced image analysing techniques like the smooth-subtract algorithm \citep[for a description, consider][]{Peissker2020b} or the Lucy-Richardson algrorithm \citep[][]{Lucy1974} resulted in the observation and detection of the short-period star S62 and S4711-S4715 \citep[][]{Peissker2020a, Peissker2020d}. While the observation of the S47xx stars suffer from close-by sources but also overall data impacts like the variable background or general weather conditions, the detection of S62 is robust and follows a clear Keplerian trajectory \citep[][]{Peissker2020a}.\newline While S2 takes about 16 years to orbit Sgr~A*, \cite{Meyer2012_s55} derived a orbital timescale for S55/S0-102 of $t_{period}\,=\,12.8\,\pm0.1$ years. Subsequently, the S-cluster star S62 orbits Sgr~A* much faster with $t_{period}\,=\,9.9\,\pm\,0.3$ years. This is outpaced by S4711 with an orbital period of $t_{period}\,=\,7.6\,\pm\,0.3$ years \citep[][]{Peissker2020d}.\newline
In addition to the hot spots\footnote{Blobs/Clouds of hot plasma.}, the S-cluster stars S2, S55, and S62 are perfect candidates to investigate the gravitational effect caused by the super massive black hole that resides in our galactic center \citep[][]{Lynden-Bell1969, Eckart2002, Gravity2019}. The importance of independent stellar probes is underlined by \cite{Tursunov2020} who investigates the electromagnetic forces that could influence plasma clouds close to Sgr~A*. Moreover, the spin of the SMBH may induce a small but measurable impact on these blobs \citep[][]{Fragione2020}. Arguably, stellar probes really close to Sgr~A* ($<\,20$ AU) may also suffer from the influence of the black hole spin but are not affected by electromagnetic forces.\newline
Therefore, the observation of S62 but also S4711 and S4714 is important to identify the gravitational phenomena  of Sgr~A* in its direct vicinity.
Since SINFONI (VLT) and NIRC2 (KECK) data of 2019 became publically available by mid 2020, we investigate these additional data sets and do a photometric analysis to compare the reported K-band magnitudes for S62 by \cite{Gillessen2009}, \cite{Peissker2020a}, and \cite{GravityCollaboration2021}. We also use mainly NACO data between 2002 and 2019 to identify the S-cluster star S29 on its Keplerian trajectory towards Sgr~A* year by year. Additionally, we analyse the orbital elements for S29 of \cite{Gillessen2009}, \cite{Gillessen2017}, and this work.}

\normalfont{In Sec. \ref{sec:data}, we introduce the used telescopes, data, and methods. This is followed by the observational results summarized in Sec. \ref{sec:results}. We then discuss the material in Sec. \ref{sec:discussion} also in comparison with the literature. Finally, we draw conclusions in Sec. \ref{sec:conclusion}. In the appendix, we list the used data and present a K-band spectrum for S62 extracted from the SINFONI data cube of 2019.}

\section{Data and Analysis} \label{sec:data}

Here, we will outline the additional data compared to \cite{Peissker2020a}. Also, we describe the data reduction and the methods used in this work. For the NIRC2 data, we used the public available observations from the Keck Observatory archive\footnote{\url{https://koa.ipac.caltech.edu/}}. The downloaded data was already reduced. 

\subsection{Data reduction}

The data was observed with SINFONI, a near-infrared (NIR) integral field spectrograph that allows long exposure times by using adaptive optics. Due to the high NIR sky variability, the usual observation scheme is object-sky-object (o-s-o). Any other observational scheme could enhance the over- and under-subtraction of emission features as shown by \cite{Davies2007}. We will comment on that in detail in an upcoming publication (Pei$\beta$ker et al., in prep.).\newline 
The exposure time of a single data cube is 600 sec. For reducing the data, we use the ESO pipeline with the usual steps like FLAT FIELD-, BAD PIXEL-, and SKY-correction. The final single data cubes are cleaned by cropping the nonlinear border. After this, they are placed and stacked in a $100\,\times\,100$ pixel array (spatial dimensions) to create the final mosaic. For comparison with the here investigated epoch of 2019.5 observed with SINFONI, we use published data by GColl (simulated image) and \cite{Do2019} (observed with the KECK telescope) both showing the GC in 2019.4. The NIRC2/KECK data were observed in the K-band with an exposure time of 2.8 sec/coadd with a pixel scale of 0.009 arcsec \normalfont{and downloaded from the Keck Observatory Archive (KOA)\footnote{Principale investigator: Tuan Do, UCLA (USA).}.\newline
Except in 2014 and 2019, we use published and deconvolved NACO data to identify S29 on its way towards Sgr~A*. Most parts of this data were reduced, analyzed, and published in \cite{muzic2010,Witzel2012, Sabha2012, Shahzamanian2016, Parsa2017, Ali2020, Peissker2020a}. However, for the data between 2014 and 2019, we redo the Lucy-Richardson deconvolution as described before. With this, we ensure an independent approach to the data compared to \cite{Peissker2020a}. Throughout this manuscript, we use the term {\it deconvolution} and {\it high-pass filter} interchangeably because they describe the same process. We are not using any other image analysis technique as the Lucy-Richardson technique.}

\subsection{High-pass filter and methods}

From the final cube, a K-band image (collapsing the spectral range between $2.0\,-\,2.2\,\mu m$) is extracted. This K-band image is placed in a $256\,\times\,256$ pixel array where we shift S2 to the central position. An artificial Point Spread Function (APSF) is created with values that are based on a Gaussian fit of S2 from the final mosaic. The advantage of a APSF is that the imperfections of the natural SINFONI-PSF (extraction over a \normalfont{finite} spatial support, noisy wings, etc.) \normalfont{are not affecting the photometry or the deconvolution in general}.\newline 
After this, we subtract several different backgrounds from the input file to reduce noise and to eliminate the chance of a false positive. We use $10^4$ iteration steps with the Lucy-Richardson deconvolution algorithm. For convolving the data, we use a PSF with a Full Width Half Maximum (FWHM) of around $70-80\%$ of the input PSF. To verify the resulting image, we cross-compare the S-star positions with known stellar positions and data, if available.

For the photometric analysis, we use as a reference the derredened K-band magnitude of S2. By comparing the peak intensity counts of the stellar source with the reference star, we derive with
\begin{equation}
    mag_{S62}\,=\,mag_{S2}\,-\,2.5\,\times\,log(ratio_{S62/S2})
\label{eq:1}    
\end{equation}
the magnitude of the source of interest.\newline 

For the orbit plots, we use a Keplerian model. \normalfont{As mentioned earlier, most} of the S29 data points are extracted from the same set that is used for the analysis of the Schwarzschild precision presented in \cite{Parsa2017}. Later, these results were independently confirmed by \cite{GravityCollaboration2020} using GRAVITY. This confirmation underlines the astrometric robustness of the imaging data used for the analysis by \cite{Parsa2017}.

\section{Results} \label{sec:results}

\normalfont{In this section,} we will present the detection of S62 in 2019.4 and 2019.5 with NIRC2 (KECK) and SINFONI (VLT), respectively. Furthermore, we \normalfont{show discrepancies of the observations carried out with the SINFONI and NIRC2 with the simulated stellar positions presented by \cite{GravityCollaboration2021}. We extract from the same publication the position of S62 and compare it with the data shown in \cite{Peissker2020a}.} To minimize the chance of confusion, we repeat the analysis and detection of S62 for the observation of the GC in 2018. Moreover, we present a timeline for S29 that shows the S-star on its trajectory towards Sgr~A* between 2002 and 2019. We apply an additional photometric analysis to the observation of S29 based on the high-pass filtered images. 

\subsection{The Keplerian orbit of S62}

\normalfont{Given the increased attention regarding the orbit of S62, we want to revisit the data presented in \cite{Peissker2020a} to check the possibility of a linear trend as claimed by, for example, \cite{GravityCollaboration2021}. If so, the derived Keplerian orbit would not represent the motion of the star. 
\begin{figure*}[ht!]
\includegraphics[width=1.0\textwidth]{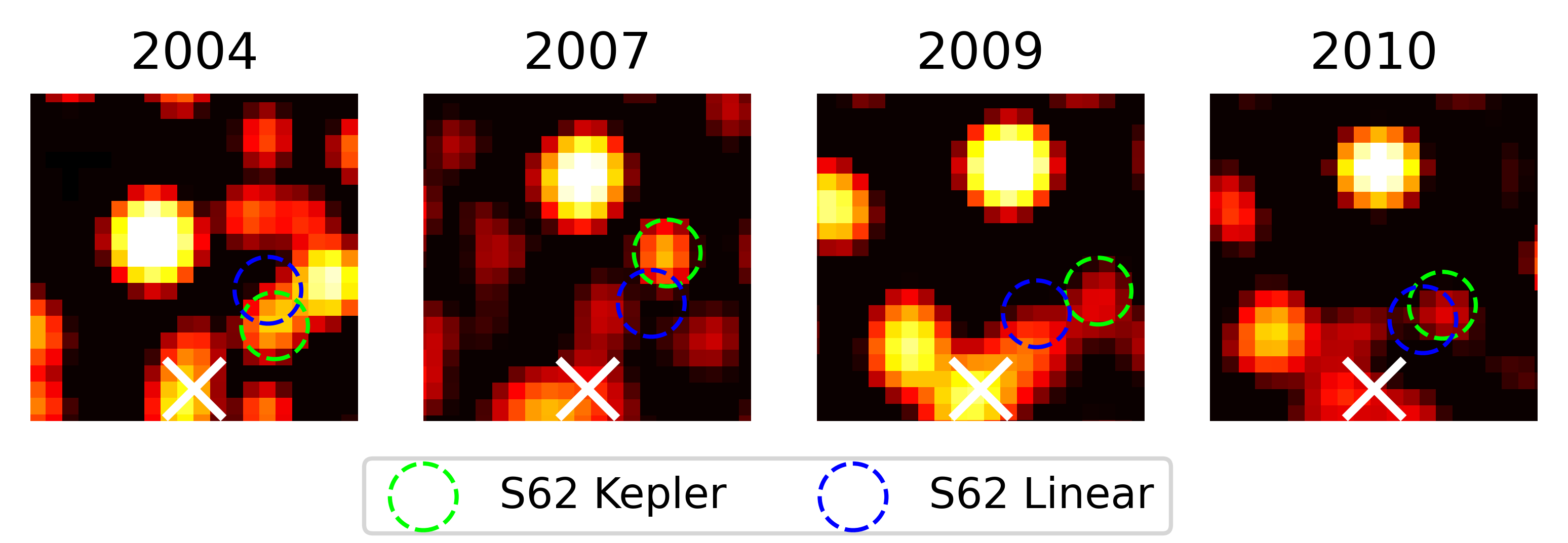}
\caption{Comparison of the position of linear S62 position (dashed lime circle) taken from GColl and the observed S62 location (dashed blue circle) with K-band NACO data of 2004, 2007, 2009, and 2010. Sgr~A* is indicated by a white $\times$. The size of the panels is about $0.2"\,\times\,0.2"$. North is up, East to the left. As implied by the images, a linear trajectory of S62 does not reflect the orbital position of the star.} 
\label{fig:4}
\end{figure*}
In Fig. \ref{fig:4}, we show the results of the analysis. For examining the position of S62, we use high-pass filtered K-band NACO images (see Sec. \ref{sec:data}).
The blue dashed circle corresponds to the data points from \cite{GravityCollaboration2021}. We extracted the R.A. and DEC. data points of 2004, 2007, 2009, and 2010 from the related publication by using the software digitizer (Thomas Ott, MPE Garching)\newline
We find that the Keplerian orbit positions, indicated with a green dashed circle, are in line with the orbit fit presented in \cite{Peissker2020a}. Furthermore, we do not see the need to classify the orbital motion of S62 as linear. Especially if we consider the clear observation of S62 in 2004 and 2007 eliminates the chance for a non-Keplerian orbital description. With a maximum uncertainty of about one pixel (i.e., 13.3 mas), we do find inside the green, and therefore Keplerian, circle a source that we identify as S62 which is in agreement with \cite{Gillessen2009}. As we show in the next subsections, the identification of S62 can been confused with nearby stellar sources (like S29) in the crowded area close to Sgr~A*.}




\subsection{Identification of S62 in 2018 and 2019}

\begin{figure*}[t!]
\includegraphics[width=1.\textwidth]{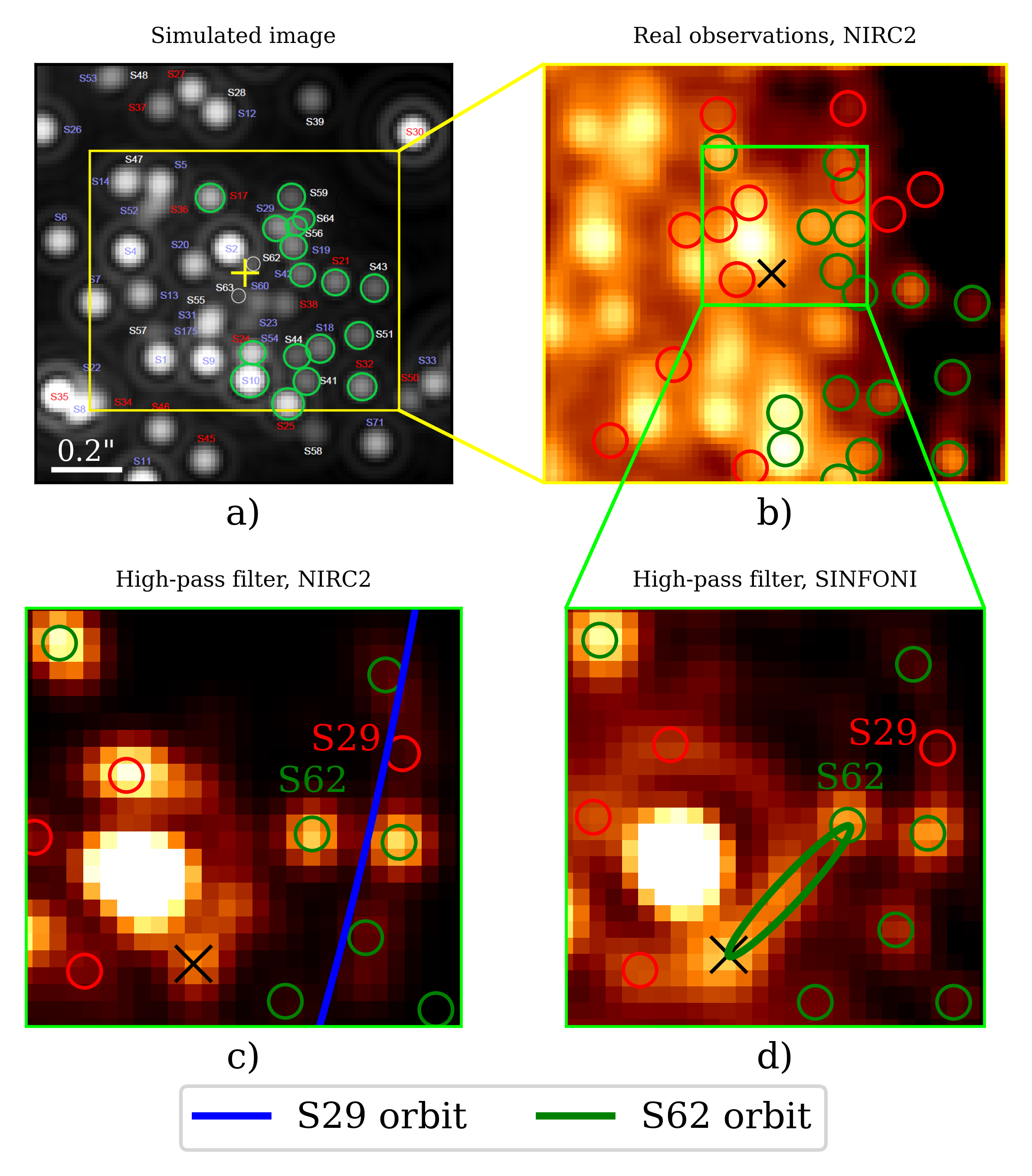}
\caption{Galactic center observed with SINFONI/VLT and NIRC2/KECK in 2019. The $\times$/+ marks the position of Sgr~A*. With red circles in panel b), c), and d) we mark stellar sources which are confused or not shown in a) (which is taken from GColl). The size of panel b) is about  $0.9"\,\times\,0.8"$, c) and d) around $0.3"\,\times\,0.3"$. In every panel, North is up, East to the left. The blue segment in panel c) is based on the orbital elements of S29 given in Tab. \ref{table1} and shown in Fig. \ref{fig:2}. The orbital elements of \cite{Peissker2020a} are used to implement the green orbit of S62 shown in panel d). \normalfont{Panel c) and d) show deconvolved K-band images. Because of cosmetic reasons, we smooth the data in panel b) with a 3 px Gaussian.}} 
\label{fig:1}
\end{figure*}

\begin{figure}[h!]
\includegraphics[width=.5\textwidth]{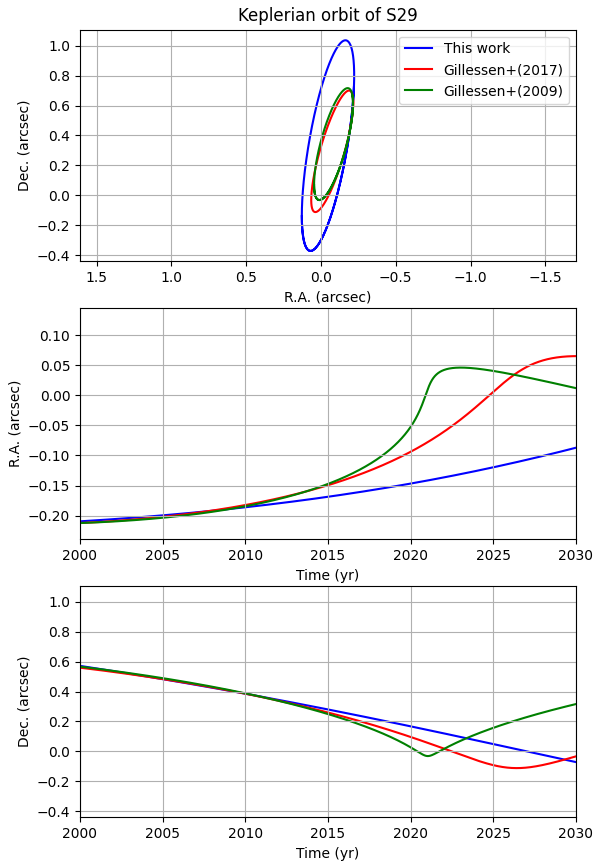}
\label{fig:2}
\caption{On-sky projected orbits of S29 based on \cite{Gillessen2009} (green), \cite{Gillessen2017} (blue), and this work (blue) based on NACO and SINFONI data between 2002 and 2019. The three orbital solutions for S29 starts to deviate in 2015. According to the fitted solutions in \cite{Gillessen2009}, \cite{Gillessen2017}, and this work, the periapse approach of S29 is dated to 2021.00 (green), 2025.96 (red), and 2047.01 (blue), respectively.} 
\end{figure}

By using the Lucy-Richardson algorithm on the SINFONI and NIRC2 data, we independently find the expected orbital position of S62 a stellar source in 2019.4 and 2019.5 (Fig. \ref{fig:1}, panel c and d)) as proposed by \cite{Peissker2020a}.
\begin{table*}[t!]
    \centering
    \begin{tabular}{cccccccc}
            \hline
            \hline
            Source & a [mpc] & e & i [$^\circ$] & $\omega$ [$^\circ$] & $\Omega$ [$^\circ$]&  $t_{closest}$ [years]& $t_{period}$ [years] \\
            \hline
            S29   & 28.65 $\pm$ 5.749 &  0.476 $\pm$ 0.180 & 101.70 $\pm$ 8.7 & 169.53 $\pm$ 82.65 & 349.79 $\pm$ 89.70 & 2047.01 $\pm$ 11.27 & 223 $\pm$ 59.5 \\  
            S29*   & 15.88 $\pm$ 13.4 &  0.916 $\pm$ 0.048 & 121.98 $\pm$ 11 & 343.25 $\pm$ 5.7 & 157.16 $\pm$ 2.5 & 2021.00 $\pm$ 18 & 91 $\pm$ 79 \\  
            S29**   & 17.12 $\pm$ 0.760 &  0.728 $\pm$ 0.052 & 105.76 $\pm$ 1.7 & 346.46 $\pm$ 5.9 & 161.91 $\pm$ 0.80 & 2025.96 $\pm$ 0.94 & 103 $\pm$ 2.0 \\  
            \hline
    \end{tabular}
    \caption{Orbital elements for S29 derived in this work. The values for S29* and S29** are taken from \cite{Gillessen2009} and \cite{Gillessen2017}, respectively. The S29 uncertainties are derived from the standard deviation of the listed values.}
    \label{table1}
\end{table*}
Close to the location of S62 on the descending part of the orbit (green circle on the green ellipse in Fig. \ref{fig:1}, panel d)), the star S29 can be found at its position as proposed by \cite{Peissker2020a} (red circle on the blue segment of the S29 orbit in Fig. \ref{fig:1}, panel c)) for the epoch 2019.4/2019.5.\newline 
As shown in Fig. \ref{fig:1}, stellar sources are marked with red (missing identification) and green (matching identification) circles. The sources that are marked with green circles can be independently identified in the NIRC2/KECK and SINFONI/VLT images resulting from observations in 2019.4 and 2019.5, respectively. As mentioned before, the SINFONI and NIRC2 data agree for these sources with the simulated image from the same epoch. 
The stellar sources which are marked with red circles in the SINFONI and NIRC2 image (Fig. \ref{fig:1}, panel b)) can not be identified in the simulated image \normalfont{taken from \cite{GravityCollaboration2021}} (Fig. \ref{fig:1}, panel a)) or do show a different arrangement.\newline 
\normalfont{Inspecting Fig. \ref{fig:1} and the individual panels in detail}, we identify over $30-40\%$ more stellar sources in the NIRC2 and SINFONI data, which increases the number of S-cluster members significantly. However, a detailed study of these sources exceeds the scope of this work but should be taken into account for future observations of the cluster.
\normalfont{Recently, the S-cluster star S29 gained an increased interest since it was believed that the re-identification of S62 after 2015 was confused. By comparing published orbital elements for S29 from the literature with this work, we find a satisfying agreement with the stellar trajectory between 2002 and 2009 (for a description, see caption Fig. \ref{fig:2}).}
\normalfont{However,} the different orbital solutions for S29 start to differ for years after 2010 (see Fig. \ref{fig:2}) \normalfont{and become significantly apparent in the following. The related orbital elements that were used for Fig. \ref{fig:2} are listed in Table \ref{table1}. Furthermore, we used the orbital elements to identify the S-cluster star S29 throughout the publically available NACO data (please consult the next subsection).}  
\normalfont{Based on the here presented data and the projected} on-sky position of S62 during the apoapsis in 2019.4, we derive a distance of 73.7 mas to S29 and 62.5 mas to S64 (that is located south of S29) by fitting a Gaussian to the data within a 0.1" radius around the related positions. 


\subsection{Identification of S29 between 2002 and 2019}

Based on the presented \normalfont{orbital elements} in \cite{Peissker2020a}, we trace the S-cluster S29 star between 2002 and \normalfont{2018 with the additional data point from this work in} 2019. Since NACO was not mounted at the VLT in 2014 and decommissioned in 2019, we use SINFONI data for these two years in the presented timeline (Fig. \ref{fig:3}).

\begin{figure}[ht!]
\includegraphics[width=0.5\textwidth]{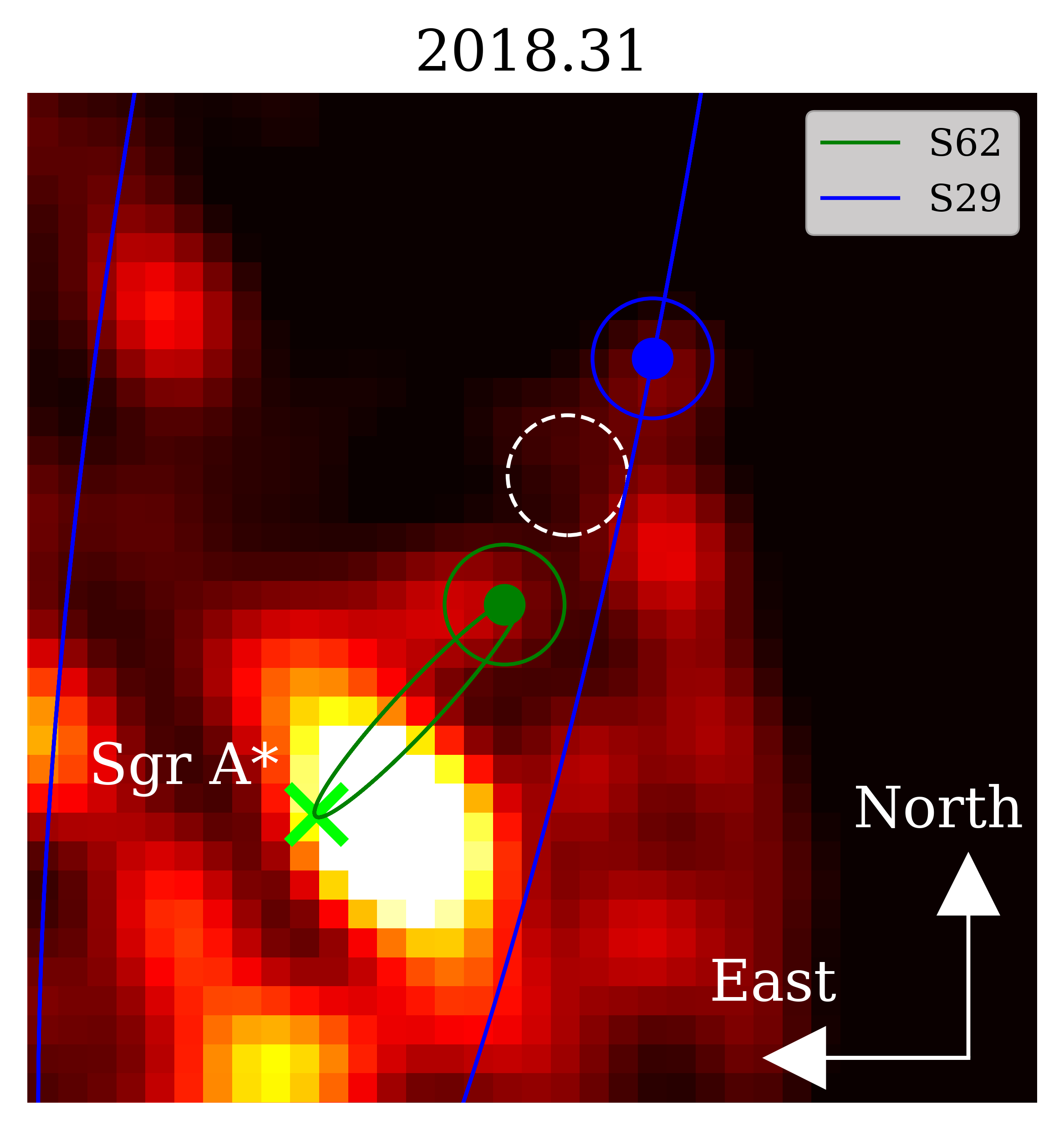}
\caption{Observation of S29 and S29 in 2018. \normalfont{In this figure}, we show high-pass filtered NACO K-band data of the position of the S-cluster members S29 (blue) and S62 (green) on their Keplerian orbit with respect to Sgr~A* (lime colored $\times$). The size of the image is $0.46''\,\times\,0.46''$, the orientation is indicated by the white arrows in the lower right corner. The white dashed circle marks the position of S29 according to \cite{Gillessen2017}.} 
\label{fig:5}
\end{figure}

\normalfont{With the increased data baseline, we repeat the analysis presented in \cite{Peissker2020a} and \cite{Ali2020}. For that}, we apply a PSF sized Gaussian (6-7 pixels) to derive the position of S29 throughout the analyzed data \normalfont{between 2002 and 2019}. The resulting positions \normalfont{from the presented data in Fig. \ref{fig:3}} are used to determine the Keplerian orbit fit, which is shown in Fig. \ref{fig:2}.\newline
We find in agreement with the orbital solution (Tab. \ref{table1}) the star S29 on the predicted orbit on its trajectory around Sgr A* (Fig. \ref{fig:3}). 
In 2014, the position of S29 can be confused with interfering close by stellar sources \normalfont{\citep[S59 and S64, see][]{Ali2020}}. In addition, the identification of S29 in 2019 can be challenging (Fig. \ref{fig:1}). Anyhow, as we show in Fig. \ref{fig:1}, Fig. \ref{fig:3}, and \cite{Peissker2020a}, we confidently minimize any confusion. Considering the well-defined orbital solution and the identification of S29, we underline the robust observation of S62 between 2002 and 2019.
\begin{figure}[t!p]
\includegraphics[width=.5\textwidth]{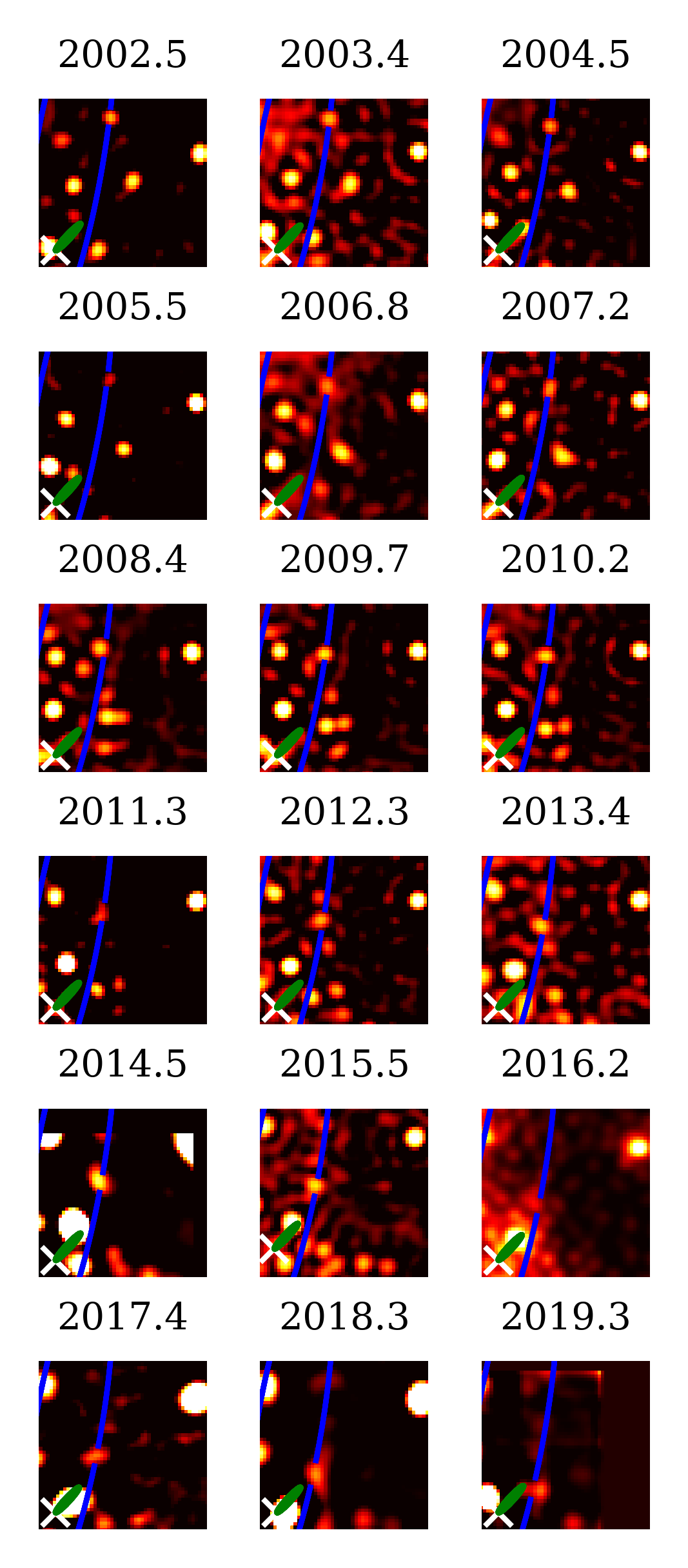}
\label{fig:3}
\caption{Timeline of the S-cluster star S29 between 2002 and 2019 \normalfont{based on the deconvolved K-band images (see the text for details)}. Expect for the SINFONI observation in 2014 and 2019, the presented data is obtained with NACO. North is up, east is to the left. The size of the panels in 2014 and 2019 is $0.62''\,\times\,0.62''$, in the remaining years $0.66''\,\times\,0.66''$.} 
\end{figure}
We emphasize the individual identification of S29 and S62 by re-analysing the K-band NACO data of 2018. In this period, S29 is sufficiently far away from Sgr~A* as we show in Fig. \ref{fig:5}. Both stars can be found at their expected positions. Tracing S29 in 2015, 2016, and 2017 agrees with this statement for the confusion-free observation in 2018. 
Considering the proposed position of S29 in \cite{Gillessen2017} for 2018.4, the location of the S-cluster star would be between S62 and S29 (see the white dashed circle in Fig. \ref{fig:5}). 
Applying a comparable large uncertainty of about $\pm\,1\,px$ to the position of S29 and S62 (represented by circles) coincides with the emission of \normalfont{the observed} point sources. \normalfont{Considering the orbital elements given in Table \ref{table1} that differs from the derived values of this work, we can not confirm a confident detection of a stellar source (see white dashed circle in Fig. \ref{fig:5}).}
\normalfont{For the distance of S29 and S62 as shown in Fig. \ref{fig:5}, we find that both stars are over 140 mas apart.}

\subsection{Photometric analysis of S29 and S62}
\label{sec:photometry}
\begin{figure}[ht!]
\includegraphics[width=.5\textwidth]{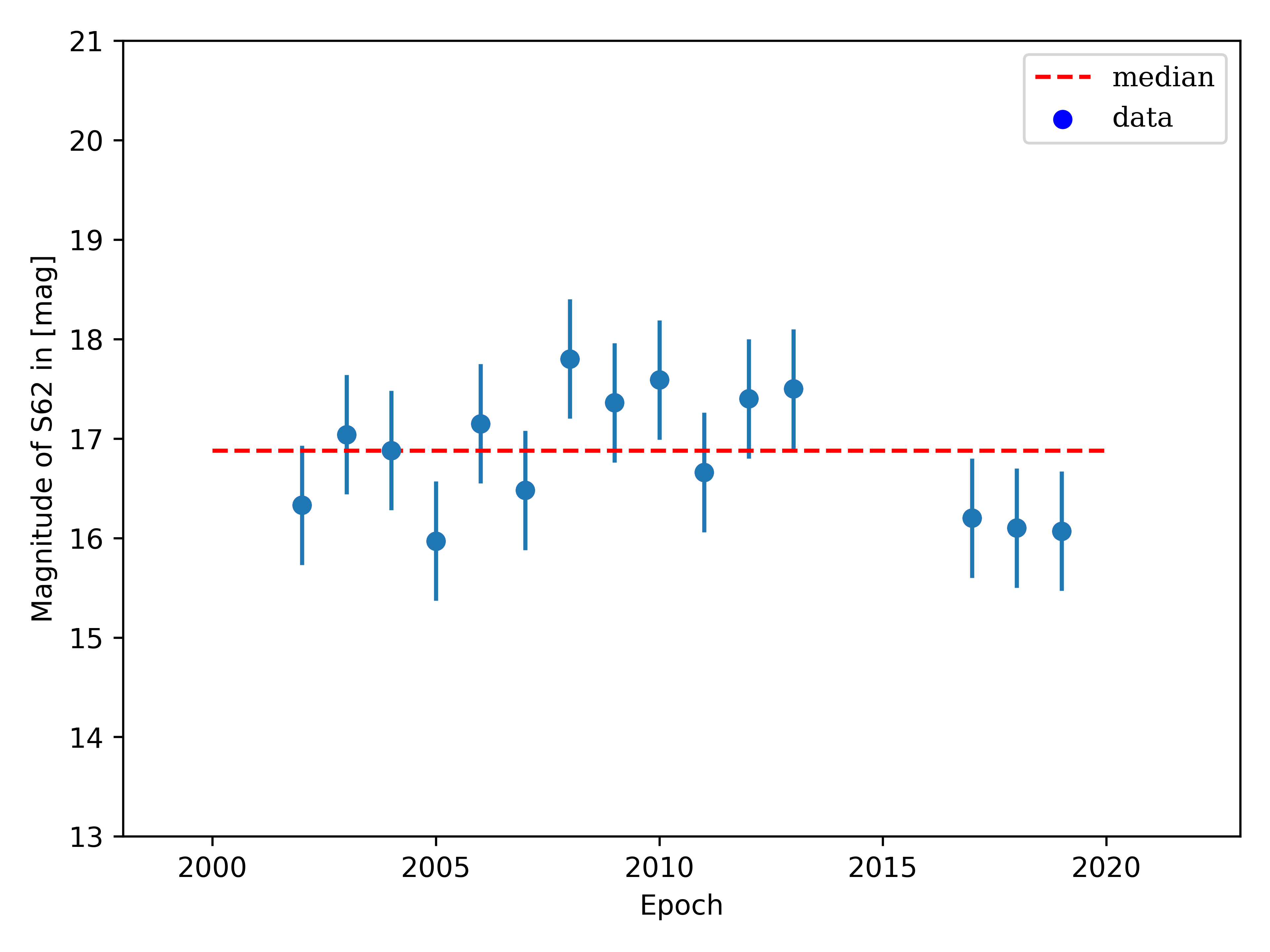}
\caption{Light curve for S62 between 2002 and 2019. The standard deviation is $0.59$ mag and hence inside the $1\,\sigma$ interval. As expected for a star in a crowded field with a varying background, the magnitude is fluctuating.} 
\label{fig:6}
\end{figure}

Even though the identification of stars in a crowded field is challenging, a photometric analysis can reduce the chance of confusion. 
\normalfont{Using} Eq. \ref{eq:1} and the high-pass filter SINFONI image of 2019.5, we derive a K-band magnitude of about 16.2 mag for S62, which is consistent with the derived value in \cite{Peissker2020a}. 
The usual uncertainty for the robust photometric analysis in such a crowded field is about $\pm\,0.2$ mag. A photometric analysis of the convolved NIRC2/KECK data shown in Fig. \ref{fig:2} delivers a K-band magnitude for S62 of $15.95\,\pm\,0.2$ mag.\newline
Additionally, we find a K-band magnitude for S62 of 16.19 mag in 2018 (Fig. \ref{fig:5}) which very well agrees with the detection in 2019 \normalfont{and underlines the robust observation of the S-star}. Expanding this analysis to the complete data set from \cite{Peissker2020a}, we derive the light curve for S62 between 2002 and 2019 (Fig. \ref{fig:6}).   
We find a median magnitude for S62 of $16.8\,\pm\,0.6$ mag and therefore matching the $1\,\sigma$ uncertainty range. \normalfont{The estimated uncertainty is based on the standard deviation.} 
Figure \ref{fig:6} demonstrates that within the uncertainties in a crowded field, S62 shows the same brightness over one entire orbital period as expected for a solid identification of a single star. We note that we already provided a detailed discussion on the difficulties in deriving brightness information on a star in a crowded field in \cite{Peissker2020a}.\newline
\normalfont{Using the data presented in Fig. \ref{fig:3}, we can confidently confirm the derived 16.7 mag for S29 as discussed in \cite{GravityCollaboration2021}. From the photometric analysis, we derive a median K-band magnitude for S29 of $16.95\,\pm\,0.25$ mag, where the uncertainty is based on the standard deviation. Compared to the S29 K-band magnitude in the literature (16.7-16.9 mag), the derived median value of this work fits well in the uncertainty range.}\newline
\begin{figure}[ht!]
\includegraphics[width=.5\textwidth]{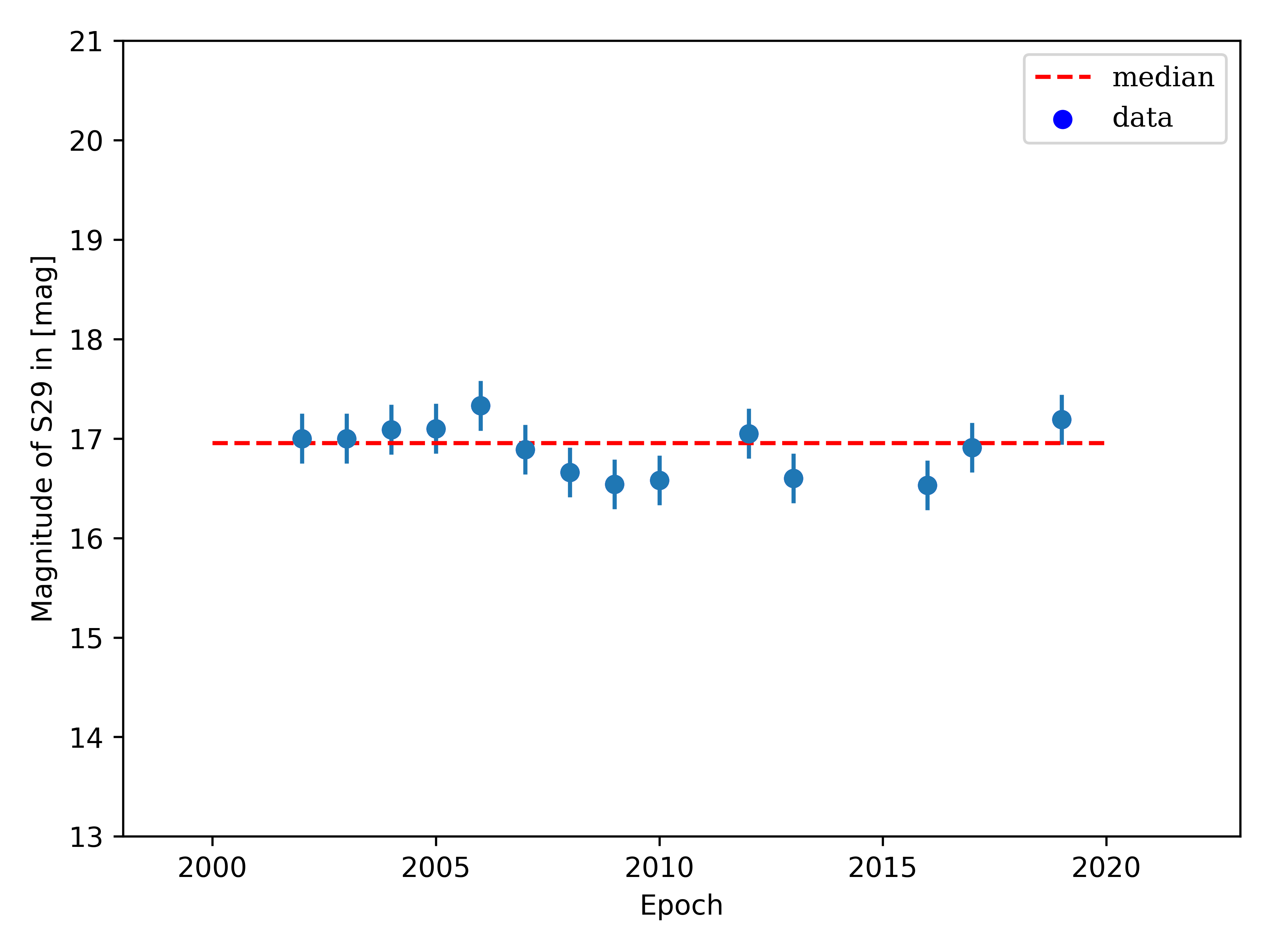}
\caption{Magnitude of S29 derived from SINFONI and NACO data. The red dashed line marks the median magnitude. The distribution of data points \normalfont{shows} that \normalfont{the K-band magnitude of} S29 is \normalfont{stable inside the 1 $\sigma$ uncertainty. We derive a standard deviation of} $0.25$ mag.} 
\label{fig:3}
\end{figure}



\section{Discussion}
\label{sec:discussion}
Here, we discuss the observation of S29 between 2002 and 2019. We will also elaborate on the clear detection of S62 in 2018 and 2019. 

\subsection{The observation of S29 and S62 between 2002 and 2019}

Additionally to the NACO data baseline from 2002 to 2018 that is presented in \cite{Peissker2020a} and \cite{Peissker2020b}, we can add the S62 detection in the SINFONI \normalfont{and NIRC2} data of 2019. As proposed by the derived orbit that is presented by \cite{Peissker2020a}, we find at the expected position a stellar source in 2019.\newline 
While all panels of Fig. \ref{fig:1} (that show SINFONI and KECK data) do agree very well in the positions and relative fluxes of stars, the simulated image of 2019.4 taken from GColl does suffer from inconsistencies. 
The most obvious inconsistencies are marked with red circles. However, the modelled stellar positions of, e.g., S44, S54, and S57 seem to be different compared to the presented KECK image. While some of the distorted positions can be explained by general crowding problems (like, e.g., a PSF overlap or blending problem and hence a variable background), missing stars are indicating an identification problem. Unfortunately, this translates into confused observations as we indicate with red circles in Fig. \ref{fig:1}. \newline
Based on the presented NIRC2 and SINFONI data, the simulated image (panel a) of Fig. \ref{fig:1}) suffers from $30-40\%$ misplaced or missing stellar sources.\newline 
Considering the photometric analysis of \cite{Gillessen2009}, \cite{Peissker2020a}, and this work, we do find at least in an arguable magnitude range matching values for S62. \normalfont{The derived S62 magnitude in this work is well inside the $1\,\sigma$ uncertainty range and independently confirms our robust detection of the S-star.\newline}
We furthermore confirm the derived magnitude for S29 by \cite{GravityCollaboration2021}. In Fig. \ref{fig:5}, we identify S29 based on the derived Keplerian orbit \normalfont{(Table \ref{table1})} and the photometric analysis at the expected position. Since the observed star is the only source in the FOV with matching properties, we conclude a robust and confusion-free detection of S29. With Fig. \ref{fig:3} and Fig. \ref{fig:5} we present data that underlines our analysis of the S-cluster members.\newline
The confusion-free observation of S29 in 2017 is moreover in agreement with presented data in \cite{Habibi2017}. The related authors present a high-pass filtered NACO K-band image as a finding chart for the epoch 2017.2. In agreement with the measured distance between Sgr~A* and S29 in the finding chart of \cite{Habibi2017} we find S29 at the expected position based on the here derived orbital solution \citep[see also][]{Peissker2020a}. Unfortunately, the predicted position by \cite{Gillessen2017} differs by about $0.06''$ from the S29 location shown in \cite{Habibi2017}.

\normalfont{Furthermore, we derive a mass of S62 of about $5.84^{+2}_{-1}\,M_{\odot}$ by following \normalfont{the empirical approach} 
\begin{equation}
    lg\frac{M_S}{M_{\odot}}\,=\,km_K\,+\,b
\label{eq:2}    
\end{equation}
where we adapt k = -0.192 and b = 3.885 from \cite{Peissker2020d}.}
\normalfont{The mass uncertainty of $1-2\,M_{\odot}$ that} is adapted from \cite{Habibi2017} since the authors use spectroscopic data for the brightest stars (magnitudes larger than 15.5 mag) in the S-cluster. This justifies the \normalfont{transference to} a fainter star like S62. This, however, would increase the \normalfont{resulting magnitude range to about $\pm\,1-2$ mag. For 2019, this would imply a magnitude for S62 of about $16.2\,\pm\,2$ mag using the presented SINFONI data. Even considering this conservative magnitude uncertainty, we can not confirm the $18.9$ mag for S62 as derived by \cite{GravityCollaboration2021}.}\newline

\section{Conclusion}
\label{sec:conclusion}
\normalfont{In this work}, we \normalfont{show} the consistent detection of S62 on its orbit around Sgr~A* in 2019 with SINFONI and NIRC2. Based on the derived orbit, mass, and magnitude from \cite{Peissker2020a}, the identification of S62 in the SINFONI and NIRC2 data is free of confusion. Considering GColl, we can not confirm their identification of S62 as a faint star with 18.9 mag, especially because of the detection limit of \normalfont{NACO. 
Furthermore,} we find the \normalfont{$\sim$17} mag S-star S29 at the expected position \normalfont{throughout the presented data covering 2002-2019.
Because} of the missing and displaced stars in the modeled and presented data of \cite{GravityCollaboration2021}, we are allowed to speculate that the authors may have confused the identification of S29 with S62. We find strong evidence that amongst the discussed candidates there is no \normalfont{star on a linear trajectory} that moves without curvature as close \normalfont{to Sgr~A*.}
The derived S62 K-band magnitude by \cite{Gillessen2009} supports our conclusion that the authors of \cite{GravityCollaboration2021} confused S62 with a 18.9 mag faint star close to \normalfont{Sgr~A*.
From this analysis, we draw some final conclusions:\newline\newline
We observe S29 and S62 at their expected Keplerian positions in 2018 and 2019. The photometric analysis of S29 and S62 underlines this finding, especially considering the clear detection in 2018 of both objects as individual stars. Even though the projected minimal distance of
S29 at the descending part of the orbit is dated to be around 2020/2021, the true periapse will happen in the next few years to come.}

\acknowledgments

We are grateful for the comments by the anonymous referee that helped to improve this work. 
This work was supported in part by the
Deutsche Forschungsgemeinschaft (DFG) via the Cologne
Bonn Graduate School (BCGS), the Max Planck Society
through the International Max Planck Research School
(IMPRS) for Astronomy and Astrophysics as well as special
funds through the University of Cologne. Conditions and Impact of Star Formation is carried out within the Collaborative Research Centre 956, sub-project [A02], funded by the Deutsche Forschungsgemeinschaft (DFG) – project ID 184018867. Part of this
work was supported by fruitful discussions with members of
the European Union funded COST Action MP0905: Black
Holes in a Violent Universe and the Czech Science Foundation
-- DFG collaboration (No.\ 19-01137J).
This research has made use of the Keck Observatory Archive (KOA), which is operated by the W. M. Keck Observatory and the NASA Exoplanet Science Institute (NExScI), under contract with the National Aeronautics and Space Administration.

\bibliographystyle{aasjournal}
\bibliography{bib.bib}

\appendix
\section{Used data}
\normalfont{In the following,} we list the SINFONI data that was observed in 2019 (Table \ref{tab:data_sinfo3}). As mentioned in Sec. \ref{sec:data}, the common observation scheme is object-sky-object (o-s-o). In contrast, some observation programs do show an unusual o-o-s-o-o pattern. Hence, we exclude all exposures that use a wrong sky due to the high K-band variability \citep[][]{Davies2007}. 
\begin{table*}[htbp!]

        \centering
        \begin{tabular}{cccccc}
        \hline\hline
        \\      Date & Observation ID &  \multicolumn{3}{c}{Amount of on source exposures} & Exp. Time \\  \cline{3-5} &   & Total & Medium & High & \\
        (YYYY:MM:DD) & &   &  &  & (s) \\ \hline 
        
        \\ 
        2019.04.20 & 0103.B-0026(B)  & 9 &   0  &  8  &    600   \\
        2019.04.28 & 0103.B-0026(B)  &  4 &   0  &  2  &    600   \\
        2019.04.29 & 0103.B-0026(B)  &  8 &   0  &  4  &    600   \\
        2019.05.02 & 0103.B-0026(F)  &  8 &   0  &  8  &    600   \\
        2019.05.23 & 5102.B-0086(Q)  &  13 &   0  &  6  &    600   \\
        2019.05.24 & 5102.B-0086(Q)  &  4 &   0  &  2  &    600   \\

        2019.06.01 & 0103.B-0026(F)  &  9 &   0  &  2  &    600   \\
        2019.06.03 & 0103.B-0026(D)  & 8 &   0  &  2  &    600   \\
        2019.06.04 & 5102.B-0086(Q)  &  6 &   0  &  3  &    600   \\
        2019.06.06 & 594.B-0498(Q)   &  11 &   0  &  10  &    600   \\
        2019.06.09 & 5102.B-0086(Q) &  14 &   0  &  10  &    600   \\
        2019.06.14 & 0103.B-0026(D)   &  4 &   0  &  0  &    600   \\
        2019.06.19 & 0103.B-0026(D) &  2 &   0  &  0  &    600   \\

        \hline  \\
        \end{tabular}
        
        \caption{SINFONI data of 2019. In total, we used 57 on-source exposures. This translates to almost 10 hours of integration time.}
        \label{tab:data_sinfo3}
        \end{table*} 
        
\section{Spectrum of S62 in 2019}

\begin{figure*}[ht!]
\includegraphics[width=1.\textwidth]{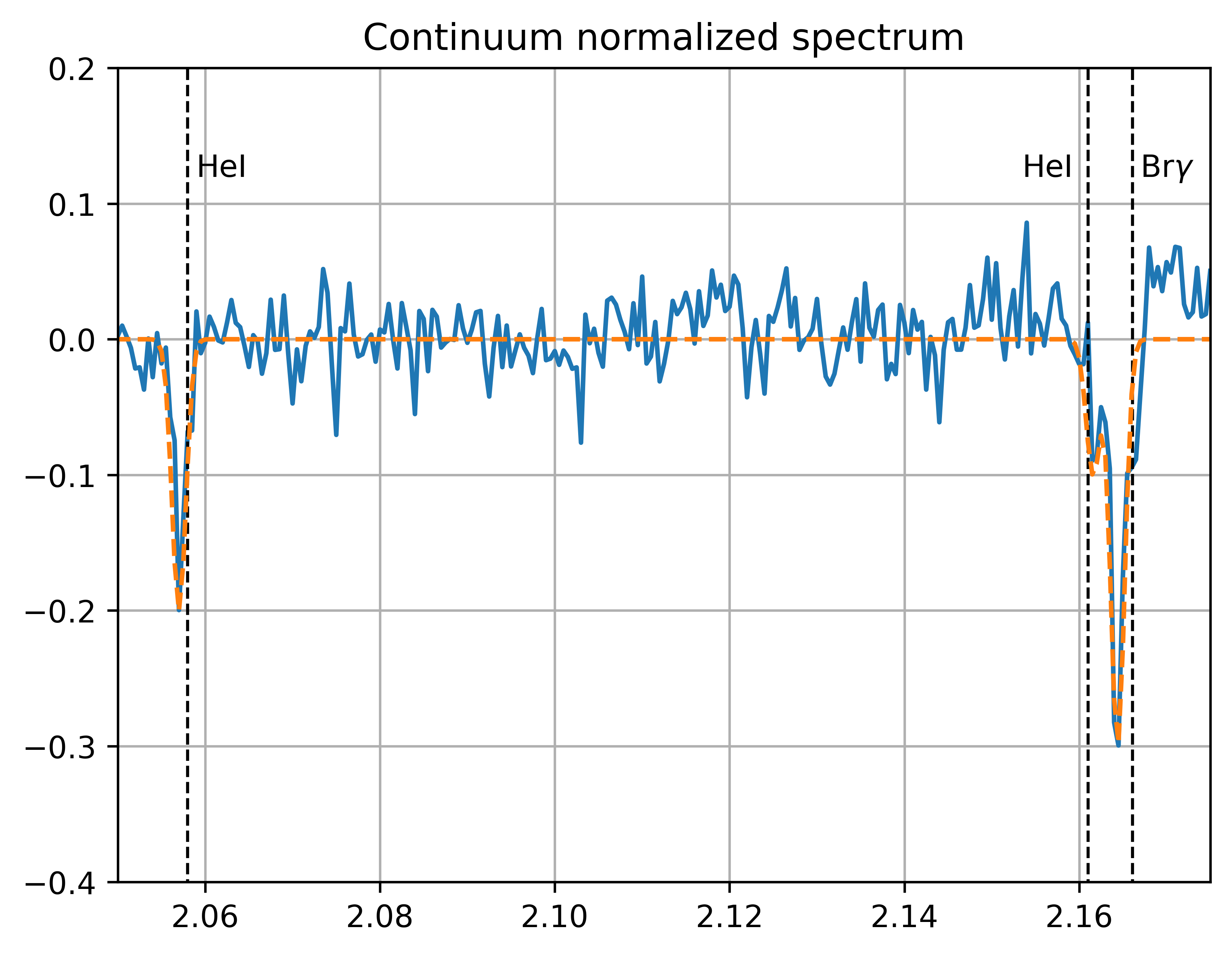}
\caption{K-band spectrum of S62 extracted from the SINFONI data cube in 2019. The blue line represents the data, the orange one the modeled fit. The y-axis is in arbitrary units.} 
\label{fig:6-appendix}
\end{figure*}

With the isolated position of S62 in 2019, we investigate the SINFONI data in 2019. We apply a 9 px sized aperture with a local background subtraction with the same dimensions. The resulting spectrum is fitted and divided by a polynomial function (Fig. \ref{fig:6-appendix}). We model 3 Gaussians to fit the absorption features at $2.057\,\mu m$ (HeI), $2.161\,\mu m$ (HeI), and $2.1644\,\mu m$ (Br$\gamma$). 
While the weak helium line at $2.161\,\mu m$ is influenced by the background and the noisy character of the data, the Br$\gamma$ line at $2.1644\,\mu m$ shows the strongest absorption. We determine the ratio between the blueshifted HeI absorption line at $2.057\,\mu m$ and the ionized HI line to be 0.66 (HeI/HI). The line of sight velocity based on the blueshifted Br$\gamma$ line is $235.44\,km/s$ and agrees very well with the derived orbit presented in \cite{Peissker2020a}. 

\section{Comment on the detection of S29 and S62}        

\normalfont{In this section, we would like to address the chance of observing an almost 19 mag faint star in the NACO, SINFONI, and NIRC2 data. Analysing the available data consistently shows S29 and S62 on their predicted and observed orbits. We have presented robust light curves of the two stars that underline this finding. Even more, independent observations are in full agreement with the analysis presented in this work and \cite{Peissker2020a}.\newline
The authors of \cite{GravityCollaboration2021} identify S62 to be a faint star with $mag_K\,=\,18.9$ mag which moves on a linear trajectory proposed by \cite{Gillessen2009} and \cite{Gillessen2017}. As we show in Fig. \ref{fig:4}, we can not confirm this statement. Given the distance and velocity of S62 to Sgr~A*, it should show some curvature, implying that a linear fit does not reflect the nature of the trajectory. 
However, we question the ability to observe such a faint star with NACO and SINFONI in the first place. Based on the published data by the authors \citep[][]{Gillessen2009, Habibi2017}, the almost 19 mag seems questionable. 
Consequently, this implies that the faint 18.9 mag star detected interferometrically with the VLTI can not be S62 because of the NACO detection limit in the crowded central arcsecond \citep[][]{Sabha2012}. However, since the sensitivity of GRAVITY is largely enhanced compared to SINFONI, NACO, and NIRC2, the chance of detecting a faint star is increased in the same way. \newline Additionally, we would like to point out that some stellar sources and positions are missing or misplaced in the finding chart of \cite{GravityCollaboration2021}. This will significantly complicate the identification of stars especially close to Sgr~A*.}

%
%
%
%
%
%
        
\end{document}